\documentstyle[prl,aps,epsfig,twocolumn]{revtex}

\begin{document}



\title{Multiplier phenomenology in random multiplicative 
cascade processes}

\author{
Bruno Jouault$^1$, 
Peter Lipa$^{2,3}$,
and Martin Greiner$^2$
}

\address{$^1$Institut f\"ur Theoretische Physik, Technische Universit\"at,
             D--01069 Dresden, Germany}
\address{$^2$Max-Planck-Institut f\"ur Physik komplexer Systeme, 
             N\"othnitzer Str.\ 38, D--01187 Dresden, Germany}
\address{$^3$Dept.\ of Radiation Oncology,
University of Arizona, Tucson AZ-85721, USA}

\date{30.4.1998}

\maketitle

\begin{abstract}
We demonstrate that the correlations observed in conditioned multiplier
distributions of the energy dissipation in fully developed turbulence can be
understood as an unavoidable artefact of the observation procedure. Taking the
latter into account, all reported properties of both unconditioned and
conditioned multiplier distributions can be reproduced by cascade models with
uncorrelated random weights if their bivariate splitting function is
non-energy conserving. For the $\alpha$-model we show that the simulated
multiplier distributions converge to a limiting form, which is very close to
the experimentally observed one. If random translations of the observation
window are accounted for, also the subtle effects found in conditioned
multiplier distributions are precisely reproduced.
\end{abstract}

\\

\narrowtext



Random multiplicative cascade processes frequently serve as phenomenological
models for the study of a variety of complex systems exhibiting multifractal
behaviour.  In particular the energy dissipation field of fully developed
 turbulent
flows has been most successfully modelled in such terms, physically motivated
by Richardson's picture of energy transfer from large to small scales by
random breakups of eddies \cite{FRI95}.

Binary multiplicative cascade models relate the energy flux $E_L$ at some
integral scale $L$ to $E_r=E_L W_1\cdots W_J$ contained in a subinterval of
size $r=2^{-J}$ at scale $J$ by a product of mutually independent random
weights $W_j$. More precisely, the energy flux $E_k^{(j)}$, contained in the
interval $k$ with length $r=L/2^j$ splits into a left (L) and right (R)
offspring interval, each of length $r/2$, whereby the content propagates
according to $E_{2k}^{(j+1)}=W_L E_k^{(j)}$ and $E_{2k+1}^{(j+1)}=W_R
E_k^{(j)}$, respectively.  For each breakup the two random weights 
$W_{L,R}\geq 0$
are chosen, {\em independently\/} from any
preceding breakup, according to a joint probability density $p(W_L,W_R)$
with expectations $\langle W_{L,R}\rangle{=}1/2$. The latter
we denote as splitting function.  Note that for a full description of a binary
breakup the specification of the {\em joint\/} density is necessary. For the
special case where the splitting function is concentrated along
the diagonal $W_L+W_R=1$, i.e.\ $p(W_L,W_R)=p(W_L)\delta(W_L+W_R-1)$, each
breakup strictly conserves energy; for all other forms, however, the relation
$E_{2k}^{(j+1)}+E_{2k+1}^{(j+1)}=E_k^{(j)}$ holds only on average.

A simple example is the splitting function
\begin{eqnarray}
\label{eq:amod}
  \lefteqn{p(W_L,W_R)   = } \\
& & 
\quad\left[p_1 \delta ( W_L - (1-\alpha)/2 ) + p_2 \delta ( W_L - (1+\beta)/2 )
       \right]* \nonumber \\
& & 
\quad\left[
       p_1 \delta ( W_R - (1-\alpha)/2 ) + p_2 \delta ( W_R - (1+\beta)/2 )
 \right], \nonumber
\end{eqnarray}
which is known as the $\alpha$-model \cite{SCH84}. The probabilities
$p_1=1-p_2=\beta/(\alpha+\beta)$ are determined by
energy conservation in the mean: $p_1(1-\alpha)+p_2(1+\beta)=1$.  In contrast
to the otherwise similar $p$-model suggested in \cite{MEN87}, the 
$\alpha$-model does not conserve energy in each local
splitting, since with probability $p_1^2+p_2^2$ we have $W_L+W_R \neq 1$.

Multiplicative cascade models are based on two assumptions: {\bf (a)} 
the existence
of a scale-independent splitting function $p(W_L,W_R)$ and {\bf (b)}
statistical independence of the random weights $W_{L,R}$ at one breakup from
those of any other breakup.  Once $p(W_L,W_R)$ is chosen or deduced from
experiment, the assumptions (a) and (b) allow to determine all moments and
scaling exponents of the energy dissipation $\varepsilon_r{=}E_r/r$ by inverse
Laplace transforms \cite{NOV71,GRE98}. Many experimental measurements of
moments and scaling exponents confirm that this simple construction reproduces
the measured multifractal aspects of the energy dissipation field amazingly
well (e.g. \cite{MEN87,MEN91}).

A series of experimental investigations \cite{CHA92,SRE95,PED96} aim at
a direct study of the random weights $W$ by analysing the distributions of
closely related observables, so-called multipliers $M$ (or breakup
coefficients $q{=}2M$ in the case of binary breakups), whose operational
definition is given below.  These studies reveal that assumption (a) is true
over a decade of scales, with the restriction that the scale-invariant
distribution $p(M)$ depends also on the relative position of the offspring to
the parent interval; however, assumption (b) is clearly violated.  Significant
correlations between multipliers at adjacent scales are observed. Such
correlations apparently obscure the validity of uncorrelated multiplicative
cascade models for fully developed turbulence, and more elaborate models, such
as the `correlated $p$-model' \cite{SRE95}, were suggested. These experimental
findings gave already rise to a critical discussion of the limitations of
multiplier phenomenology \cite{NEL96}.  In this Letter we present a further
clarification of the experimental results.

In order to explain the observed multiplier correlations within the framework
of binary multiplicative cascade processes two considerations are made:
{\bf(i)} the bivariate splitting function $p(W_L,W_R)$ should not be assumed
to be energy conserving; this is justified by the experimental restriction to
measure the energy dissipation field of three-dimensional turbulence only
along a one-dimensional cut (i.e. from the velocity time series obtained from
anemometers and employment of Taylor's frozen flow hypotheses).  Consideration
{\bf(ii)} concerns the obvious non-homogeneity of cascade processes.  Due to
their hierarchical structure the $n$-point correlation functions of the
generated energy dissipation field are not translationally invariant
(cf.\cite{GRE97} for a visualisation); multipliers in dyadic intervals will be
different if the observation window $L$ is shifted by an arbitrary amount, say
$\Delta x$.  In real world experiments, however, the observation window is
placed in no relation with the unobservable hierarchical structure of the
cascade, so that implicitly an averaging over uniform random translations is
performed. The effect of such random translations was hitherto assumed to be
negligible. We test this assumption by introduction of random shifts before
calculating the multipliers.  It is especially this latter operation that
leads to a full explanation of all available experimental findings of
conditional multiplier distributions.  In the following we discuss the
implications of above considerations (i) and (ii) one after the other.

It has already been pointed out in \cite{GRE96} that the non-conservation of
energy in the splitting function leads to deviations from perfect multifractal
scaling. Here we focus on its influence on multiplier distributions obtained
from simulated realizations $E_k^{(J)}$ after $J$ cascade steps.  Depending
on the relative position of parent to offspring intervals, left, right and
centred multipliers at scale $j$ and position $k$ are operationally defined
as
\begin{equation}
\label{eq:mult}
M_{k,L}^{(j)}=
\frac{\bar{E}_{2k}^{(j+1)}} {\bar{E}_{k}^{(j)}}, 
\quad 
M_{k,R}^{(j)}=
\frac{\bar{E}_{2k+1}^{(j+1)}} {\bar{E}_{k}^{(j)},} 
\end{equation}
\[ 
M_{k,C}^{(j)}=
\frac{\bar{E}_{4k+1}^{(j+2)} + \bar{E}_{4k+2}^{(j+2)}} 
{\bar{E}_{k}^{(j)}},
\]
and the multiplier distributions $p(M^{(j)})$ are obtained by histogramming
all $0 \leq k < 2^j$ possible multipliers at a given scale.

It is important to note that the experimentally measured `backward' energies
\begin{equation}
\label{eq:backeps}
\bar{E}_{k}^{(j)}=\sum_{l= 
k 2^{J{-}j}}^{(k+1) 2^{J-j}-1}
E_{l}^{(J)},
\end{equation}
are obtained by successive summations from the finest resolution scale $J$ to
larger ones and are generally not equal to the `forward' energy densities
$E_{k}^{(j)}$, which arise as intermediate states in the evolution of the
cascade from larger to smaller scales.  For a clearer distinction we denote
the former with a bar.  Since, by definition, the multipliers satisfy
$M_{k,L}+M_{k,R}=1$, they effectively enforce local energy conservation and
therefore give only incomplete information on the true parent-offspring
relation of non-energy conserving breakups.

For the special case of energy-conserving splitting functions, such as the
$p$-model \cite{MEN87}, the multipliers (\ref{eq:mult}) do give a faithful
representation of local breakups.  Here, the left and right multiplier
distributions are equal to the splitting function $p(W)$ along the $W_L+W_R=1$
diagonal and do not depend on the scale $j$.

The situation is quite different for non-energy conserving models, such as the
$\alpha$-model. In this case there is no simple analytical relation between
the splitting function (\ref{eq:amod}) and the resulting distributions of
multipliers (\ref{eq:mult}); in fact, the mapping from (\ref{eq:amod}) to
$p(M^{(j)})$ via Eqs.\ (\ref{eq:mult}) and (\ref{eq:backeps}) acts effectively
as a non-linear smoothing operation. This we demonstrate in the following
simulation results: The scale dependent left\footnote{In all simulations the
distributions for left (L) and right (R) multipliers are identical; therefore
only one (L) is shown.  Differences occur, however, with centred multipliers
and the latter are shown in Figs.\ 2 and 3.} multiplier distribution
$p(M_L^{(j)})$ is depicted in Fig.\ 1, where $10^5$ configurations were
generated, each with $J=9$ cascade steps.  While the weights $W_{L,R}$ in the
forward evolution may take only two values $(1-\alpha)/2$ and $(1+\beta)/2$,
the multipliers (\ref{eq:mult}) take more and more distinct values as the
scale difference $J-j$ increases and becomes soon quasi-continuous.  Already
after three backward steps the left (and right) multiplier distributions
apparently converge to a limiting form which comes surprisingly close to a
parameterisation of the experimentally observed multiplier distribution
\cite{SRE95}, shown as continuous line. The parameters used in this simulation
were $\alpha=0.3, \beta=0.65$, but this finding also holds for other
parameters as, for example, $\alpha=\beta=0.4$ or $\alpha=0.5,
\beta=0.3$; all these choices have approximately the same second order
splitting moment $\langle W_L^2\rangle=\int_0^1 p(W_L,W_R)\, W_L^2\, dW_L
dW_R$.

A similar convergence follows for centred multiplier distributions $p(M_C)$
with the distinction that the corresponding limiting density appears to be
narrower (shown in Fig.2a for $j=3$) compared to $p(M_L)$; this latter feature
is quite in agreement with the observations in \cite{PED96}.  Analogous
results are found for multipliers generalised to arbitrary length ratios
between offspring and parent intervals. In Fig.\ 2a the left-skewed histograms
show corresponding limiting distributions for left and central multipliers for
two cascade steps $j\rightarrow j+2$ and thus relate intervals with length
ratio 1/4.

In above comparisons the unavoidable random translations of the observation
window with respect to the cascade position (as discussed in suggestion (ii)
above) has not been accounted for.  Following the procedure introduced in
Ref.\cite{GRE97}, we adopt a scheme of random translations in the following
way: for the target resolution scale $J$ of a given integral length scale $L$
a longer cascade realization with $J+3$ steps is generated (corresponding to
an integral length scale $8L$) and an observation window of size $L$ is
shifted randomly by $t$ bins within $8L$. In other words, only the $2^J$ bins
of the generated $E^{(J+3)}_{k'}$, which lie within the randomly placed
observation window are considered, giving a simulated and translated
$E^{(J)}_k=E^{(J+3)}_{k'-t}$, where $t$ is a uniformly distributed integer
within $[0,\,7{\cdot} 2^J{-}1]$.  Then the multipliers are determined again by
(\ref{eq:mult}) and sampled over a large number of configurations and random
shifts $t$.

There is no dramatic change in unconditioned multiplier distributions after
addition of random translations; the small effect is illustrated in Fig.\ 2b
compared to Fig.\ 2a.  The left multiplier distribution $p(M_L)$ now is even
closer to the experimentally deduced scale-invariant parametrisation of
Ref.\cite{SRE95}; this property also holds for the other two parametrisations
used, $(\alpha,\beta)=(0.5,0.3)$ and $(0.4,0.4)$.  Note, that the centred
multiplier distribution $p(M_C)$ now suffers a small but noticeable asymmetry
which is also seen in experimental observations \cite{PED96}.  Moreover, the
gaps close to the endpoints at $M=0$ and $M=1$ are now filled, which is a
point of interest in the discussion of Novikov's `gap-theorem'
\cite{PED96,NEL96,NOV94} and weakens any conclusions drawn from 
it\footnote{The input splitting function of the $\alpha$-model has gaps
$(1-\alpha)/2$ at $W{=}0$ and $(1-\beta)/2$ at $W{=}1$; both gaps disappear in a
multiplier analysis.}.

So far the multiplier distributions we have looked at are unconditioned.  In
the experimental analyses \cite{SRE95,PED96} conditioned multiplier
distributions of the form
\begin{equation}
\label{eq:conddist}
p\left(M^{(j)}_\Delta \left| M_{\text{min}} \leq M^{(j-1)}_\Delta 
\leq M_{\text{max}}\right.\right)
\end{equation}
have been shown; they correlate a parent multiplier $M^{(j-1)}_\Delta$ with
the multiplier $M^{(j)}_\Delta$ of its offspring, where $\Delta$ stands for
$L,C,R$ respectively.

The addition of random translations introduces significant correlations among
multipliers at different scales, which are well reflected in the conditioned
distributions (\ref{eq:conddist}). Figs.\ 3a and 3b illustrate the
correlations between offspring and parent multiplier for the centred and left
case respectively.  Compared to the unconditioned density $p(M^{(j)}_C)$ the
conditioned one $p(M^{(j)}_C| 0 \leq M^{(j-1)}_C \leq 1/2)$ is skewed to the
left and $p(M^{(j)}_C| 1/2 < M^{(j-1)}_C \leq 1)$ is skewed to the right; the
centred offspring and parent multipliers are positively correlated. Moreover,
the maximum value of the former is higher than the later.  This result is
identical to the experimental finding of Ref.\cite{PED96}.

For the conditioned left multiplier distributions $p(M^{(j)}_L| M_{\text{min}}
\leq M^{(j-1)}_L \leq M_{\text{max}})$ we show results in Fig.\ 3b; again the
parameters $(\alpha,\beta)=(0.3,0.65)$ for the $\alpha$-model have been
used. The conditioning on the sub-range $[M_{\text{min}},M_{\text{max}}]=
[0.2,0.4]$ leads to narrowing of the distribution and conditioning on
$[M_{\text{min}},M_{\text{max}}]=[0.6,0.8]$ to a broadening.  Again, this is
in perfect agreement with the experimental finding in Ref. \cite{SRE95}.

We report without figures that for the parameter choice
$(\alpha,\beta)=(0.4,0.4)$ the effects seen in Fig.\ 3b become weaker and
vanish almost completely for the choice $(\alpha,\beta)=(0.5,0.3)$. Also Fig.\
3a is modified: for $(\alpha,\beta)=(0.5,0.3)$ the maximum of the left skewed
distribution (1) is lower than the one of (2), while for
$(\alpha,\beta)=(0.4,0.4)$ they are about equal. A noticeable left/right
shift, however, remains in all three cases.

From these observations we come to the following conclusion: with non-energy
conserving splitting functions and the inclusion of random translations the
unconditioned multiplier distributions observed in the data \cite{SRE95,PED96}
are reproduced quite naturally; moreover, once the input splitting function is
skewed in a certain direction also the correct conditioned multiplier
distributions are deduced. Very similar numerical observations are also
obtained for different choices of splitting functions \cite{JOU98}. Since the
simulated cascades satisfy both assumptions (a) and (b) above, we regard the
comparable violation of (b) in experiments and Fig.\ 3 as an unavoidable
artefact of the observation procedure.

Related to the multiplier phenomenology is the problem of how to extract the
correct (multifractal) scaling exponents. The findings of Ref.\cite{GRE96}
favour the left/right over the central multipliers, but there the random
translations were not accounted for. Certainly the inclusion of the latter
modifies the scaling exponents to some extent \cite{GRE97}, the more noticeable
the higher the order of the underlying moments.  Due to these operational
ambiguities we feel that it should not be a matter of how to extract scaling
exponents, but how to deduce the optimal non-energy conserving and skewed
splitting function from data. A starting point are the findings of Ref.\
\cite{GRE98}, which show how to extract bivariate splitting functions;
however, the aspect of translation invariance has not been taken into account.
A satisfactory solution of this intricate inverse problem will only be
feasible once additional observables are studied experimentally, such as
$n$-point correlation functions or their wavelet compressed form \cite{GRE95}.

\acknowledgements
B.J.\ acknowledges support from the Alexander-von-Humboldt Stiftung. P.L.\ is
grateful to the hospitality and support of the Max-Planck-Institut f\"ur
Physik komplexer Systeme.


\newpage
\widetext
\onecolumn

\begin{figure}
\begin{centering}
\epsfig{file=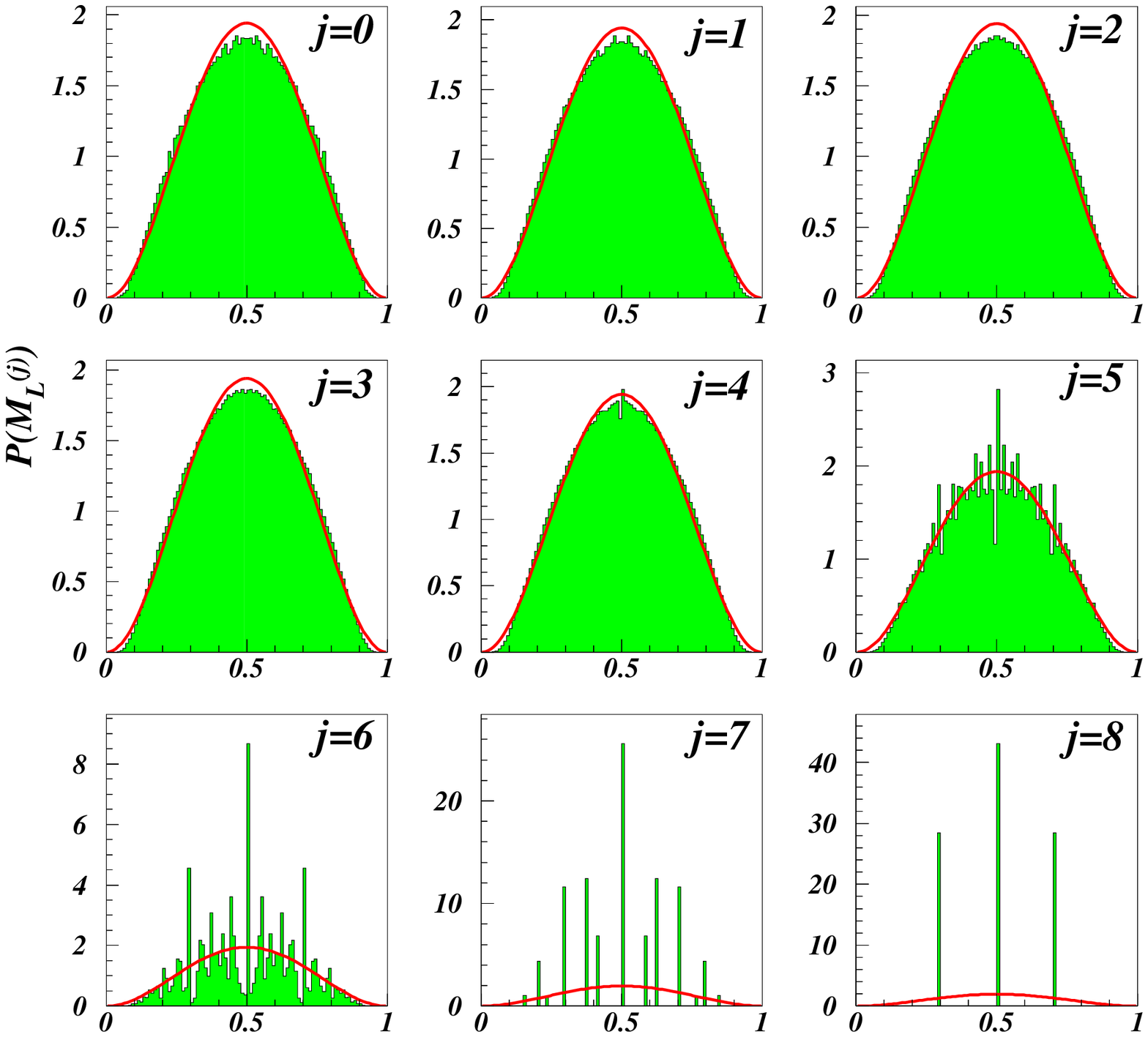,width=18cm}
\caption{Convergence of the left multiplier distributions 
$p(M_L^{(j)})$ of the 
non-energy conserving $\alpha$-model ({\protect\ref{eq:amod}}) to a 
quasi-continuous limiting form.
Parameters are $J=9$, 
$\alpha=0.3$ and $\beta=0.65$. For comparison the experimentally deduced 
Beta-function parametrisation of the left (right) unconditioned 
multiplier distribution with 
$\beta=3.2$ {\protect\cite{SRE95}} is shown as a solid line.} 
\end{centering}
\end{figure}

\begin{figure}
\begin{centering}
\epsfig{file=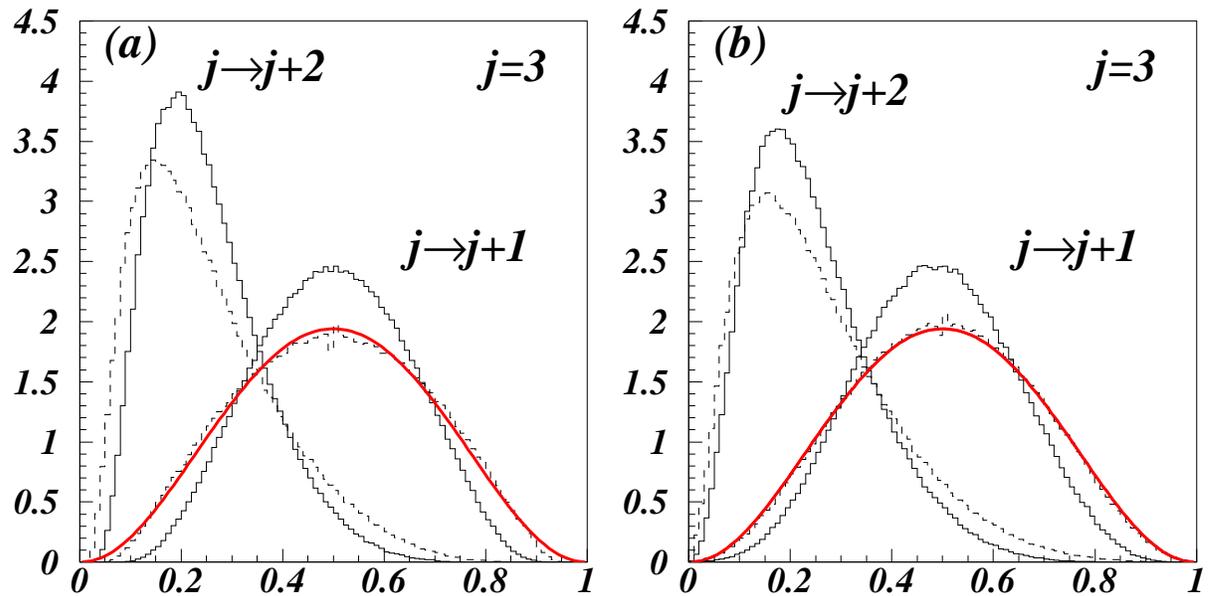,width=18cm}
\end{centering}
\caption{(a) Left (dashed) and centred (solid) 
multiplier distributions for the $\alpha$-model ({\protect\ref{eq:amod}}).
For the one step multipliers $(j \rightarrow j+1)$ the length ratio between
offspring and parent intervals is $\lambda=\frac{1}{2}$ while for the two-step
multipliers $(j \rightarrow j+2)$ we have $\lambda=\frac{1}{4}$. Parameters
are $J=9$, $\alpha=0.3$ and $\beta=0.65$.(b) Same as (a), but with random
translations applied before calculating the multipliers.  For comparison the
experimentally deduced Beta-function parametrisation of the left (right)
multiplier distribution $(j \rightarrow j+1)$ with $\beta=3.2$
{\protect\cite{SRE95}} is shown as a thick solid line.}
\end{figure}

\begin{figure}
\begin{centering}
\epsfig{file=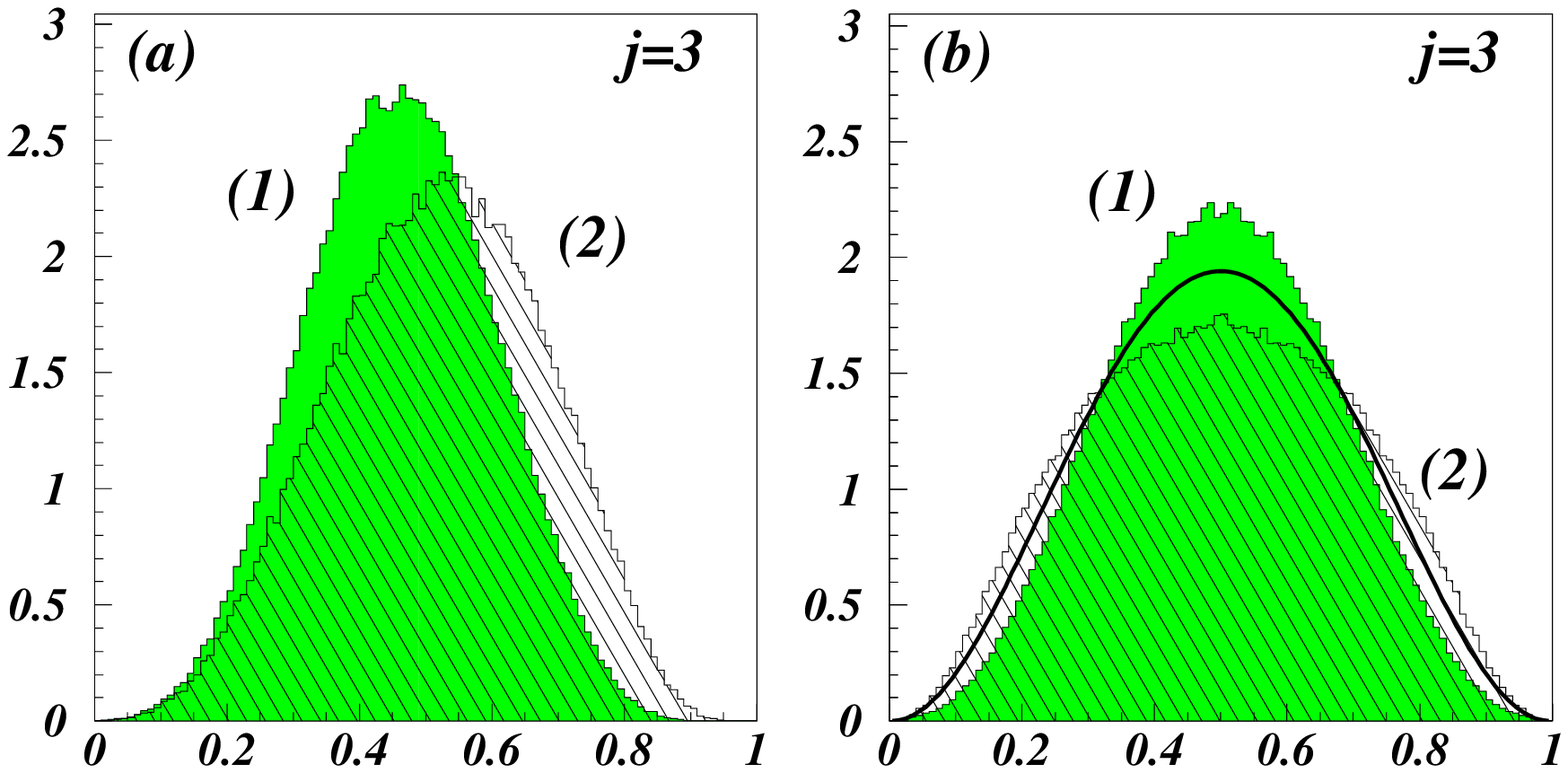,width=18cm}
\end{centering}
\caption{Conditioned multiplier distributions for the $\alpha$-model, 
satisfying strictly conditions (a) and (b) in the text.  Parameters are
$J=9$, $\alpha=0.3$, $\beta=0.65$  and random translations were applied before
calculating the multipliers. 
(a) Central distributions conditioned on central parent multipliers:  
(1) $p(M^{(j)}_C| 0 \leq M^{(j-1)}_C \leq 1/2)$
and 
(2) $p(M^{(j)}_C| 1/2 < M^{(j-1)}_C \leq 1)$.
(b) Left distributions conditioned on left parent multipliers:  
(1) $p(M^{(j)}_L| 0.2 \leq M^{(j-1)}_L \leq 0.4)$
and 
(2) $p(M^{(j)}_L| 0.6 \leq M^{(j-1)}_L \leq 0.8)$
compared to the Beta-function parametrisation of unconditioned
distributions in Fig.\ 1 and Fig.\ 2 (full line).}
\end{figure}


\begin{thebibliography}{99}

\bibitem{FRI95}
  U.\ Frisch, {\em Turbulence} 
  (Cambridge University Press, Cambridge, 1995).

\bibitem{SCH84}D.\ Schertzer and S.\ Lovejoy, 
        in {\it Turbulent Shear Flows 4}, University of Karlsruhe (1983), 
        edited by L.J.S.\ Bradbury and al (Springer, Berlin,1984).         

\bibitem{MEN87}C.\ Meneveau and K.R.\ Sreenivasan,
         Phys.\ Rev.\ Lett.\ {\bf 59}, 1424 (1987).

\bibitem{NOV71}
  E.\ A.\ Novikov,
  Appl.\ Math.\ Mech.\ {\bf 35}, 231 (1971).

\bibitem{GRE98}
M.\ Greiner, H.\ C.\ Eggers and P.\ Lipa, preprint
chao-dyn/9804024; M.\ Greiner, J.\ Schmiegel, F.\ Eickemeyer, P.\ Lipa 
and  H.\ C.\ Eggers, preprint chao-dyn/9804028, Phys.Rev. E, in press.

\bibitem{MEN91}
         C.\ Meneveau and K.R.\ Sreenivasan,
         J.\ Fluid Mech.\ {\bf 224}, 429 (1991). 

\bibitem{CHA92}
  A.\ B.\ Chhabra and K.\ R.\ Sreenivasan,
  Phys.\ Rev.\ Lett.\ {\bf 68}, 2762 (1992)

\bibitem{SRE95}
  K.R.\ Sreenivasan and G.\ Stolovitzky,
  J.\ Stat.\ Phys.\ {\bf 78}, 311 (1995).

  
\bibitem{PED96}
  G.\ Pedrizzetti, E.A.\ Novikov and A.A.\ Praskovsky,
  Phys.\ Rev.\ E{\bf 53}, 475 (1996).

\bibitem{NEL96}M.\ Nelkin and G.\ Stolovitzky,
        Phys.\ Rev.\ E{\bf 55}, 5100 (1996).

\bibitem{GRE97}
  M.\ Greiner, J.\ Giesemann, and P.\ Lipa, 
  Phys.\ Rev.\ E {\bf 56}, 4263 (1997).

\bibitem{GRE96}
   M.\ Greiner, J.\ Giesemann, P.\ Lipa and P.\ Carruthers,
        Z.\ Phys.\ C {\bf 69}, 305 (1996).
  
\bibitem{NOV94}
  E.\ A.\ Novikov,
  Phys.\ Rev.\ E{\bf 51}, R3303 (1994).

\bibitem{JOU98}
  B.\ Jouault, P.\ Lipa and M.\ Greiner, in preparation.
  
\bibitem{GRE95}M.\ Greiner, P.\ Lipa and P.\ Carruthers,
        Phys.\ Rev.\ E {\bf 51}, 1948 (1995). 
        

        

\end{thebibliography}
\end{document}